\documentclass[graybox]{svmult}

\usepackage{bm}		% boldmath
\usepackage{mathptmx}       % selects Times Roman as basic font
\usepackage{helvet}         % selects Helvetica as sans-serif font
\usepackage{courier}        % selects Courier as typewriter font
\usepackage{type1cm}        % activate if the above 3 fonts are
\usepackage{makeidx}         % allows index generation
\usepackage{graphicx}        % standard LaTeX graphics tool
\usepackage{multicol}        % used for the two-column index
\usepackage[bottom]{footmisc}% places footnotes at page bottom
\usepackage{amsmath}
\usepackage{amssymb}
\usepackage{mathrsfs}
\usepackage{amsfonts}
\newcommand{\s}{\sigma}

\newcommand{\T}{\tau}

\newcommand{\w}{\omega}
\newcommand{\aid}{a^{\dagger}_{i}}
\newcommand{\ai}{a^{\phantom{\dagger}}_{i}}
\newcommand{\fsd}{f^{\dagger}_{\s}}
\newcommand{\fsp}{f^{\phantom{\dagger}}_{\s'}}

\makeindex             % used for the subject index
\begin{document}

\title*{Quantum phase transition in the spin boson model}
\author{S. Florens, D. Venturelli and R. Narayanan}
\institute{S. Florens \at Institut N\'{e}el, CNRS and UJF, 38042 Grenoble, France
\and D. Venturelli \at Institut N\'{e}el, CNRS and UJF, 38042 Grenoble, France 
\and R. Narayanan \at Department of Physics, Indian Institute of Technology, Chennai-600036, India }
\maketitle

\abstract{In this paper we give a general introduction to quantum critical
phenomena, which we practically illustrate by a detailed study of the low 
energy properties of the spin boson model (SBM), describing the dynamics 
of a spin $1/2$ impurity (or more generically a two-level system) 
coupled to a bath of independent harmonic oscillators. 
We show that the behavior of the model is very sensitive to the bath 
spectrum, in particular how the properties of the quantum critical point in the SBM 
are affected by the functional form of the bath Density of States (DoS). 
To this effect, we review the renormalization group (RG) treatment of the SBM for 
various bath DoS, based on an unconventional Majorana representation of the spin $1/2$ 
degree of freedom. We also discuss the derivation of Shiba's relation for the
sub-ohmic SBM, and explicitely derive an effective action vindicating the quantum to 
classical mapping.}

\section{Introduction}
\label{sec:intro}
 Quantum Phase Transitions (QPT) have recently become a widespread topic in the
realm of modern condensed matter physics. QPT are phase transformations that
occur at the absolute zero of temperature and are triggered by varying a temperature 
independent control parameter like pressure, doping concentration or magnetic field.
There are various examples of systems showing quantum critical behavior, which include 
the anti-ferromagnetic transition in heavy fermion material like 
$\rm{CeCu}_{\rm{6-x}}\rm{Au}_{\rm{x}}$, that is brought about by changing the
$\rm{Au}$ doping \cite{rmpvojta}. Another prototypical example of a system
exhibiting quantum critical behavior is the Quantum Hall Effect, wherein a
two-dimensional electron gas is tuned, via an externally applied magnetic field, through 
a quantum critical point (QCP) that intervenes between two quantized Hall plateaux. 
Other examples of QPT include the ferromagnetic transition in metallic magnets as a function
of applied pressure, and the superconducting transition in thin films. 
 
Since there are such a wide range of experimentally accessible systems that
show quantum critical behavior, it is imperative that we understand QPT
at a fundamental level. We shall here endeavor to do just 
so by giving an introductory account of this fascinating phenomenon. 
As a striking illustration, we will be comparing and contrasting QPT with the 
case of more usual thermal (classical) phase transitions (as will be seen later 
on, thermal phase transitions are also referred to as classical transition,
since quantum fluctuations become unimportant in their vicinity). Let us first
begin by discussing the ferromagnetic transition, in order to better illustrate 
the rich phenomenology of phase transitions (both classical and quantum).
The route that we take here to understand the fundamentals of QPTs is as
follows: We shall first review the basic phenomenology of classical (thermal)
phase transitions. Then, we shall illustrate via heuristic arguments how
quantum fluctuations can be disregarded in the vicinity of a thermal
phase transition. These arguments also provide clues to the domain in the phase
diagram where one expects quantum fluctuations to dominate. Also, we shall briefly
discuss the question of observability of QPTs. Finally, we shall end with a
discussion of the so called quantum to classical mapping.

Let us first start with classical (thermal) phase transitions.
As a physical system, say a ferromagnet, approaches its ordering, there is a length
scale called the correlation length, $\xi$ that diverges in a power-law fashion
when one comes closer to the critical point, $\xi \sim |t|^{-\nu}$, so that the
system becomes progressively self-similar. Here, $t$ is
dimensionless parameter characterizing the distance to criticality, $t = 
\frac{T-T_c}{T_c}$ (for a thermal transition), and $T_c$ is the critical 
temperature where the phase transformation occurs.
Now, the above divergence of the correlation length encapsulates the information
that the fluctuations of the order-parameter (say the magnetization) become spatially
long-ranged as the system approaches the critical point. Analogous to $\xi$ one
can define a time scale $\xi_{\tau}$, that also diverges as a power law as one
approaches a second order transition. Thus, we have:
\begin{equation}
\xi_{\tau} = \xi^{z} = |t|^{-\nu z}
\label{se:eq:intro1}
\end{equation}
The quantity $z$ that controls the divergence of $\xi_{\tau}$ is the so called
dynamical exponent. Now, associated with this time scale we can define a
frequency scale $\omega_c \propto 1/\xi_{\tau}$, and through it a
corresponding energy $\hbar \omega_c$, which encodes information pertaining 
to the energy scale related with order-parameter fluctuations. 
This quantity $\hbar\omega_c$ competes with $k_B T_c$, the typical energy 
associated to thermal fluctuations. Now, the question of importance of 
quantum fluctuations can be re-cast into a query of which among these two energy
scales prevails. Since $\omega_c \rightarrow 0$ as one approaches the critical 
point, the energy scale of thermal fluctuations (for any non-zero $T_c$) always 
dominates over the scale $\hbar\omega_c$. In other words for a transition that 
happens at a finite temperature, $\hbar\omega_c \ll k_B T_c$. Thus, it can be 
argued that asymptotically close to a finite temperature transition, it is the 
thermal fluctuations that are the driving mechanism. This irrelevance of quantum 
fluctuations near a thermal phase transition is the reason to why they are given 
the moniker ``classical phase transition''. 

Now, from our discussion in the previous paragraph it is but obvious that if the
transition were to occur at $T=0$ (tuned by a non-thermal parameter like doping
or pressure), then the fluctuations that will drive the transition will be
wholly quantum mechanical in origin. 
It is obvious that one then needs to apply ideas from quantum statistical 
mechanics to understand QPT, as pionneered by Hertz in a seminal paper \cite{hertz} 
to tackle the problem of quantum criticality in itinerant magnetic systems. 
By using this case of the quantum magnet as a test-bed example, Hertz \cite{hertz} 
showed that any generic $d$ dimensional quantum system can be mapped onto an 
equivalent $d+z$ dimensional classical model. This statement is referred to as 
the quantum to classical mapping and is of fundamental importance in the field of QPTs. 
By using the quantum to classical mapping one can show that the critical behavior of the
quantum model is equivalent to that of a classical model but in $z$ higher
dimensions. Although this mapping is believed to be robust for insulating
magnets, it was however later shown \cite{belitz} that Hertz's conclusions
were erroneous for a large class of itinerant QPT.
This break-down in fact occurs due to the presence of soft modes in the systems 
(e.g. the particle-hole excitations in itinerant magnets) other than the 
order-parameter modes. The presence of these modes induces an effective long-ranged 
interaction between the order-parameter modes, thereby altering the critical 
behavior \cite{belitz}, as compared to Hertz's original results.

Since QPT occur at zero temperature, it was initially thought that the study
of these phase transitions was a mere academic exercise. However, it was soon
realized that the presence of a zero-temperature critical point (practically
inaccessible) can actually influence the behavior of the system at {\it finite}
temperatures. In other words, at any finite temperature, the critical singularities 
associated with the QCP are cut by the temperature, so that one observes non-trivial 
temperature dependence of various observables in a so-called quantum critical regime. The
calculation of the quantum critical regime for various models is well beyond the
scope of this work. However the interested reader is directed towards the
following papers investigating the effect of non-zero temperatures on QPT in
magnetic systems \cite{millis,sachdev}. Also, one will refer to Sec.~\ref{subsec:sbm2} 
for a brief description of the quantum critical regime in the spin boson model (SBM), 
the specific model of interest in this manuscript.

Thus, from the discussion of previous paragraphs, it is obvious that QPT
are an extremely interesting physical phenomenon to study. As alluded to before, in
this manuscript we choose to study a specific toy model example, namely the QPT 
encountered in the spin boson model (SBM), a variant of Caldeira-Leggett type models 
\cite{legget}, wherein a quantum particle is subjected to an external dissipative environment.
In the case of the SBM, this quantum ``particle'' is essentially a two-level
system, such as a spin $1/2$ impurity. While there have been many studies on dissipative
quantum models that focused on the effect of decoherence on intermediate time
scales, their behavior in the long time limit in the presence of quantum
critical points remain relatively un-explored. However, the study of such
regimes is extremely important as anomalous low energy properties emerge due to
quantum critical modes. In other words, due to the presence of a QCP, the SBM
can display non-trivial dynamics at very long times.
 
As a more general remark, we note that this model can also be used to study quantum 
criticality at the level of a single spin $\frac{1}{2}$ impurity embedded in a 
correlated system such as Mott insulators~\cite{shraiman_giam,novais,vojta_bura}, or 
magnetic metals~\cite{larkin,loh}). It also appears as an effective theory for bulk 
materials themselves (e.g. in quantum spin glasses~\cite{georges}, heavy fermion 
compounds~\cite{sengupta, si}), via the framework of DMFT. 

The SBM is introduced in Sec.~\ref{sec:sbm}. The various phases of the SBM and
the possibility of a QPT between them is discussed in Sec.~\ref{subsec:sbm2}. In
Sec.~\ref{sec:maj}, we re-write the SBM by using a Majorana fermion
representation for the impurity spins. Sec.~\ref{sec:rg} is devoted to the
derivation of the RG equations by using the Majorana representation presented
in Sec.~\ref{sec:maj}. In Sec.~\ref{sec:flow}, we look at the consequences of
the flow equation derived in Sec.~\ref{sec:maj}. Sec.~\ref{sec:lri} is dedicated
to the quantum/classical mapping of the SBM to the long-ranged Ising model. Sec.~\ref{sec:ward} is
concerned with the development of a special identity in the SBM model that is
used in Sec.~\ref{sec:shiba} to derive the so-called Shiba's
relation in the case of the sub-ohmic spin boson model. Sec.~\ref{sec:qucl}
is dedicated to a discussion on the status of the quantum to
classical mapping in the SBM, that we use as a conclusion and future outlook
regarding quantum phase transitions in dissipative models.

\section{The Spin Boson Model}
\label{sec:sbm}
\begin{figure}
\centering
{\includegraphics[height=6cm,width=7cm]{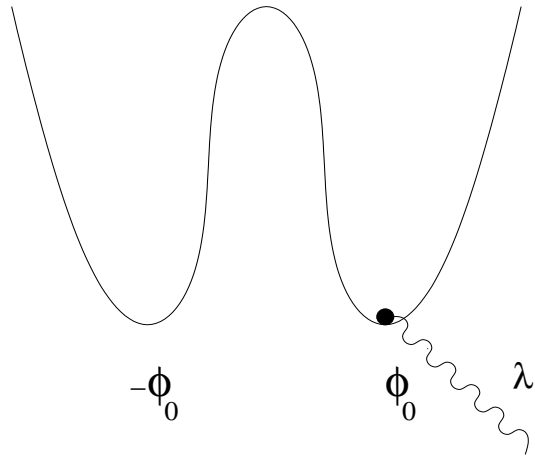}\qquad
\caption{ Fig.~\ref{fig:se:SBM} is a pictorial sketch of the SBM. It
represents a two level system (e.g. a particle in two potential minima $\pm
\phi_0$, or a spin $1/2$ impurity, coupled to a bath of harmonic oscillators
(such as phonons, nuclear spins,...) via a coupling constant $\lambda$ (wavy 
line in this figure).}
%\label{fig:se:v1}
\label{fig:se:SBM}}
\end{figure} 

As stated earlier, the SBM describes the effect of an external dissipative
environment on the quantum mechanical evolution of a two-level system. 
We will introduce in Sec.~\ref{subsec:sbm1} the general properties of the SBM,
and present in Sec.~\ref{subsec:sbm2} its possible phase diagram, obtained on 
heuristic grounds via an analysis of the various limiting cases.
\subsection{The Model}
\label{subsec:sbm1}
 The SBM involves a single spin $\frac{1}{2}$ impurity $\vec{S}$, interacting
with a set of bosonic bath variables, $a_i$, and $a_i^{\dagger}$ (in second
quantization). The interaction
between the bath's oscillator displacement and the spin is controlled via a 
coupling constant $\lambda$. Thus, the SBM Hamiltonian has the general functional form:
\begin{equation}
H = - \Delta S^x + \epsilon S^z + \lambda S^z \sum_i (\aid+\ai) +
\sum_i \w_i \aid \ai.
\label{eq:1.1}
\end{equation}
Here, in Eq.~\ref{eq:1.1}, $\Delta$ and $\epsilon$ are the transverse and
longitudinal magnetic fields respectively, applied to the quantum spin. A
physical sketch for such a SBM, wherein a two-level impurity (when the bias 
field $\epsilon$ is set to zero) is connected to an external environment,
is depicted in Fig.~\ref{fig:se:SBM}.
All that remains to completely specify the model is to endow the bosonic degrees
of freedom with a spectrum. This bosonic density of states (DoS) is taken here 
to be continuous and power-law like, and conforms to the functional form:
\begin{equation}
\rho(\w) \equiv \sum_i \delta(\w-\w_i) = 
\frac{(s+1)\w^s}{\Lambda^{1+s}}\theta(\w)\theta(\Lambda-\w).
\label{eq:1.2}
\end{equation}
Here, in Eq.~\ref{eq:1.2}, $\Lambda$ is a high-energy cutoff. When the exponent
$s$ is such that $0<s<1$, then the model is said to be in the sub-ohmic
regime, the case $s=1$ is referred to as ohmic, while the case $s>1$ is called
super-ohmic. In fact, as will be seen in the course of this paper,
the quantum critical behavior of the SBM is crucially dependent on the exponent
$s$ controlling the behavior of the bath spectrum. 

It is also convenient to define a new bosonic variable corresponding to the 
``local'' displacement:
\begin{equation}
 \phi \equiv 
\sum_i (a_i + a_i^{\dagger}),
\label{eq:phi_defn}
\end{equation}
which has an associated DoS given by,
\begin{equation}
\rho_{\phi}(\w) = -\frac{(s+1)|\w|^s}{\Lambda^{1+s}} {\rm{sgn}}(\w)\theta(\Lambda^2-\w^2).
\label{eq:1.2b}
\end{equation}
To make comparision with existing literature, one can alternatively characterize 
the bath by means of a spectral function:
\begin{equation}
J(\w) \equiv \sum_i \pi \lambda^2 \delta(\w-\w_i) = 
2\pi\alpha\w^s\Lambda^{1-s}\theta(\w)\theta(\Lambda-\w).
\label{eq:1.2c}
\end{equation}
Here $\alpha$ is the non-dimensional dissipation strength defined as $\alpha =
(s+1)\frac{\lambda^2}{\Lambda^2}$. 

\subsection{The Phases of the SBM}
\label{subsec:sbm2} 

\begin{figure}
\centering
{\includegraphics[height=6cm,width=7cm]{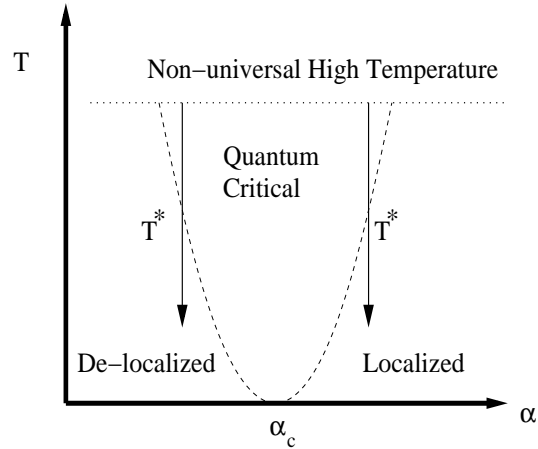}}\qquad
\caption{ Fig.~\ref{fig:se:phase_diag} shows the generic phase diagram of the
SBM model. Here, a quantum critical point $\alpha_c$ separates localized and 
delocalized phases, from which a quantum critical region emerges at finite
temperature. The scale $T^*$ is the cross-over temperature below
which various physical observables revert from quantum critical behavior to
those associated with the localized or the delocalized phase.}
%\label{fig:se:v1}
\label{fig:se:phase_diag}
\end{figure} 

As promised in Sec.~\ref{sec:sbm}, we will study here the possible phase diagram
of the SBM by looking at situations where either one of the two parameters
$\lambda$ or $\Delta$ dominates. For instance, let us first consider the case
where the dissipative coupling $\lambda$ is set to zero. The SBM then becomes
equivalent to the case of an isolated spin in a transverse magnetic field. It is well known
that such a system displays Rabi oscillations. That is, if one were to start
with an initial state pointing ``up'' along the $z$ direction, then the transverse 
field $\Delta$ periodically drives the system between up and down
configurations. This limiting case $\lambda=0$ is in fact adiabatically related to a whole 
phase at non-zero $\lambda$, dubbed for obvious reasons the delocalized 
phase, where coherent spin oscillations are expected to occur (at least for
small enough $\lambda$). We note that the average $\left<S^x\right>$
is always non-zero as long as the transverse field $\Delta$ is finite, and
thus cannot play the role of an order parameter. However, we can pursue a magnetic
analogy by noting that the longitudinal spin average $\left<S^z\right>$ is
identically zero in the delocalized phase, so that we can really relate this portion
of the phase diagram to a ground state with zero magnetization.
In fact in the alternative regime, i.e. when the dissipation $\lambda$ dominates 
over the transverse field $\Delta$, the ground state becomes doubly degenerate,
as can be checked on the trivial limiting case $\Delta=0$, where the $z$ spin
component is clearly conserved within the Hamiltonian~(\ref{eq:1.2}). A simple
physical picture emerges, with the system localizing in  one of the two minima 
at $\pm \phi_0$, see Fig.~\ref{fig:se:SBM}. Assuming adiabaticity by switching
on the transverse field, we arrive to the so-called localized phase, wherein the 
spontaneous magnetization $\big<S^z\big>\sim \big<\phi\big>\equiv M \neq0$. 
%The existence of the localized phase can be
%easily understood by laying recourse to extreme case where $\Delta$ in
%Eq.~\ref{eq:1.1} is set to zero. Then the SBM model reduces to an independent
%boson model. The independent bosonic model can be easily diagonalized by the so
%called polaronic transformation, \cite{mahan}. Now, in this situation, it can
%be shown that the probability of flipping a spin rapidly decreases with time,
%\cite{davide}. This in turn implies that the dissipation rapidly localizes the
%spin in one of the two degenerate ground states. 
%
Now, so far by using heuristic arguments, we have shown that the SBM allows for
the existence of a de-localized phase, with $\big<\phi\big> = 0$, and a localized
phase, with $\big<\phi\big>\neq0$. Thus, it is quite plausible that a
second-order phase transition takes place between the two phases. As we shall
see explicitely in Sec.~\ref{sec:flow}, there is indeed a second order quantum
localization/delocalization transition for all $0 <s \le 1$. 

The generic phase diagram for the SBM is shown in Fig.~\ref{fig:se:phase_diag},
where a $T=0$ phase transition, separating localized and delocalized
phases, takes place at a critical value $\alpha_c$ of the adimensional 
dissipation strength. 
The interesting quantum critical regime emerges above the critical point at
finite temperature, where anomalous behavior of all physical quantities is
expected.
For instance, the longitudinal spin susceptibility in the quantum critical regime 
obeys the behavior $\chi_z(T)\sim 1/T^s$, as opposed to the conventional
$1/T$ Curie-law expected for the whole localized phase. 
This anomalous power-law behavior is a direct signature of the QCP at $\alpha_c$,
and will be demonstrated in the following sections.
We note that for the ohmic $s=1$ case, the conventional treatment for studying the 
QPT is to map the SBM into an anisotropic Kondo model (AKM), and use previous
knowledge on the scaling properties of this well-known Hamiltonian.
However, in this paper, we shall follow a less well-trodden path, namely, performing 
a renormalization group (RG) calculation directly within the SBM, using a spin
represention in terms of Majorana fermions (see Sec. \ref{sec:maj} for
further details). This formalism has the advantage that it can be easily
adapted to perform calculations in the sub-ohmic limit, i.e. ($0<s<1$), see
Sec.~\ref{sec:rg}.

\section{The SBM using the Majorana representation}
\label{sec:maj}
In Sec.~\ref{sec:rg} we aim to derive the RG equations for the SBM, based on a
perturbative analysis around the localized limit, i.e. $\Delta = 0$. A technical
difficulty on this path comes from the fact that the quantum spin $\frac{1}{2}$ 
impurity does not follow either bosonic or fermionic commutation relations. 
Thus, standard calculations based on Wick's theorem cannot be invoked.
One of the many ways to avoid this problem is to map the spin-$\frac{1}{2}$
operator onto fermionic degrees of freedom, which can be done in particular using
so-called Majorana fermions. The use of this mapping to condensed matter is relatively 
recent and for a more detailed explaination the readers are referred to the following 
references, Refs.~\cite{shnir, mao}. 

The mapping between the spin $\frac{1}{2}$ impurity and the Majorana fermions
obey the following correspondence principle:
\begin{equation}
\overrightarrow{S} = -\frac{i}{2} \overrightarrow{\eta} \times \overrightarrow{\eta} 
\label{eq:maj_def}
\end{equation}
Here, in Eq.~\ref{eq:maj_def} the $\eta$ fields represent a triplet of Majorana
(real) fermions $\eta \equiv (\eta_1, \eta_2, \eta_3)$. that satisfy the
following anticommutation relations $\{\eta_i, \eta_j\} = \delta_{ij}$. 
Now, in addition to these Majoranas defined above, one can construct another
fermionic field, $\Phi = 2 i \eta_1\eta_2\eta_3$, that commutes with the Hamiltonian,
and constitutes hence a conserved quantity, with the constraint $\Phi^2 =\frac{1}{2}$. 
A very useful relation for describing the spin dynamics is given by the correspondence 
(see \cite{mao}):
\begin{equation}
 \overrightarrow{\eta} = 2 \Phi\overrightarrow{S}
\label{eq:alt_maj}
\end{equation}

Now, in terms of the Majorana fermions (see Eq.~\ref{eq:maj_def}), and
after the redefinitions $\{S_x\to S_3,S_y\to S_2,S_z\to S_1\}$ which amounts
to a $\pi/2$ rotation around the $y$ direction, the Hamiltonian of the SBM can 
be expressed as
\begin{equation}
H= - i\frac{\Delta}{2}\left(\eta_1\eta_2- \eta_2\eta_1\right)+H_B-i\lambda\phi\eta_2\eta_3
\label{eq:mh}
\end{equation}
Now in what follows, we will use Eq.~\ref{eq:mh} to perform the perturbative RG 
analysis. Before we go on to do so, a word of caution regarding the fermionic mapping 
is in order: Any mapping of the spin $\frac{1}{2}$ impurity to fermionic operators 
tends to enlarge the dimensionality of the Hilbert space. How such an enlargement is 
obviated in the case of the Majorana representation is technical matter that
goes well beyond the scope of this manuscript, and the reader is directed to 
Refs.~\cite{mao, shnir} for further details. 

\section{Perturbative renormalization group in the localized regime}
\label{sec:rg}

 \begin{figure}
\centering
{\includegraphics[height=1.2cm,width=7cm]{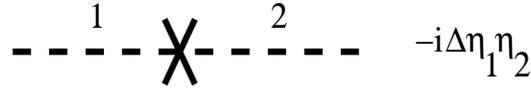}}\qquad
\caption{ Pictorial representation of the $\Delta$ vertex.}
%\label{fig:se:v1}
\label{fig:se:cross}
\end{figure} 

In this section, we will perform a perturbative RG treatment starting from the
localized phase, i.e. from the limit in which $\Delta=0$. 
Our plan of action to derive the RG equations is as follows: We initially start 
with a model of free spin ($\Delta=0$, $\alpha =0$), and then perform a perturbative 
analysis in both $\Delta$ and $\alpha$, leading to renormalizations of the 
dissipation $\alpha$ and the transverse field $\Delta$, which depend explicitely
on a generic cut-off scale $\Lambda$.
Following the philosophy of the RG, one aim to compute the renormalized
parameters at a lower cut-off scale, $\Lambda^{\prime}$, leading to so-called
flow equations.

\begin{figure}
\centering
{\includegraphics[height=3cm,width=9cm]{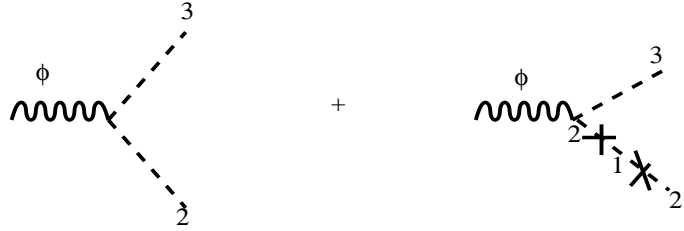}}\qquad
\caption{ Lowest order diagrams involved in the renormalization of the dissipative 
term $\lambda$.}
\label{fig:se:v1}
\end{figure}

The key ingredient in developing the perturbation theory are the free Majorana fermion
propagator $G_\eta^{\rm free}$, as well as the $\Delta$ vertex shown in 
Fig.~\ref{fig:se:cross}, and the $\lambda$ vertex (first diagram appearing in 
Fig.~\ref{fig:se:v1}).
The free fermion propagator in Matsubara frequency $\omega_n=(2n+1)\pi T$ at finite
temperature $T$ reads 
$G_\eta^{\rm free}(i\omega_n) = 1/i\omega_n$.
One can then first construct the vertex function $\Gamma_{\alpha}$ related to
the dissipative coupling $\lambda$, shown in Fig.~\ref{fig:se:v1}. 
The functional form of the vertex function can easily be deduced to be
\begin{equation}
\Gamma_{\alpha}(\omega,\Lambda) = \frac{\lambda}{\Lambda}+\frac{\lambda}{\Lambda}\Delta^2
G^{\rm free}_\eta(\omega)^2
\label{eq:rg1}
\end{equation}
Here, once again $\Lambda$ is cut-off scale, set e.g. by temperature or
the bandwidth of the bosonic modes, and $\omega$ is a frequency. The above
equation can be effectively re-written in terms of an adimensional
transverse field $h=2\Delta/\Lambda$, so that the renormalized
dissipation reads:
\begin{equation}
\Gamma(\omega,\Lambda) = \frac{\lambda}{\Lambda}\left[1-\frac{h^2}{4}\exp\left(2\ln\frac{\Lambda}{\omega}\right)\right]
\label{eq:rg2}
\end{equation}
Now, as stated in the introductory part of this section, we re-scale the cut-off
$\Lambda$ to $\Lambda^{\prime} = \Lambda -d\Lambda$. Under such a re-scaling the
vertex function can be re-written as:
\begin{equation}
\Gamma(\omega,{\Lambda}^{\prime} ) = \frac{\lambda}{\Lambda^{\prime}}\left(1+\frac{d\Lambda}{
\Lambda}\right)\left[1-\frac{h^2}{4}\exp\left(2\ln\frac{\Lambda}{\omega}\right)+\frac{h^2}{2}\frac{d\Lambda}{\Lambda}\right]
\label{eq:rg2a}
\end{equation}
The above equation can be re-written in the form of the usual RG 
$\beta$-function by including the frequency dependent vertex function into the
redefinition of the coupling constant. Then in terms of the logarithmic
differential $d\ell = -\frac{d\Lambda}{\Lambda}$, the Eq.~\ref{eq:rg2a} can be 
re-cast into the form:
 \begin{equation}
 \frac{d\lambda}{d\ell} = -\frac{\lambda}{2}h^2
 \label{eq:rg3}
 \end{equation}
and finally more compactly expressed in terms of the dimensionless
dissipation $\alpha = 2\lambda^2/\Lambda^2$ as
 \begin{equation}
 \frac{d\alpha}{d\ell} = -\alpha h^2
 \label{eq:rg4}
 \end{equation}
\begin{figure}
\centering
{\includegraphics[height=1.5cm,width=9cm]{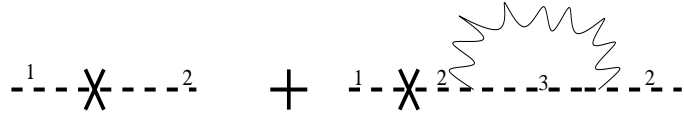}}\qquad
\caption{ Diagrams that are involved in the renormalization of the 
transverse field $\Delta$. }
%\label{fig:se:v1}
\label{fig:se:Deltacorr}
\end{figure} 

Now, in a similar vein one can calculate the first corrections to the
transverse field $\Delta$, with the diagrams depicted in Fig.~\ref{fig:se:Deltacorr}.
%which now involve the propagator of the bosonic field, hence an explicit
%dependence in the bath exponent $s$.
The technical details of this calculation are very similar to the above
calculation, and the final flow equation, written in terms of the scaled magnetic 
field $h = \Delta/\Lambda$, reads:
 \begin{equation}
 \frac{d h}{d\ell} =  (1-\alpha) h.
 \label{eq:rg5}
 \end{equation}
 
 The RG equations that we have so far derived are for the case of the ohmic damping. 
 The derivation of the flow equations in the 
non-ohmic limit is quite straight forward and can be performed by following 
the technical details elucidated above. 
Thus, for the sake of brevity we will not perform these computations here. Instead, we will just quote the results of such an exercise. 
In the presence of non-ohmic dissipation the $\alpha$ flow equation of Eq.~\ref{eq:rg4} gets modified into
 \begin{equation}
 \frac{d\alpha}{d\ell} = -\alpha h^2 + (1-s)\alpha.
 \label{eq:rg6}
 \end{equation}
However, the flow of the magnetic field $h$ retains its functional form given in Eq.~\ref{eq:rg5}, even in the presence of non-ohmic dissipation.

\section{Analyzing the RG flow}
\label{sec:flow}
In this section we shall discuss the RG flow equations that were derived above.
In Sec.~\ref{subsec:ohmic}, we shall first analyze the $\beta$ functions for 
the ohmic case ($s = 1$). Then, in Sec.~\ref{subsec:super}, we shall show that 
the super-ohmic case ($s>1$) is bereft of any critical points. 
Finally in Sec.~\ref{subsec:sub}, we shall analyze the critical behavior when 
the bath spectrum is sub-ohmic in character ($0<s<1$). 
 
 \subsection{The RG equations for the ohmic case ($s=1$)}
 \label{subsec:ohmic}
 \begin{figure}
\centering
{\includegraphics[height=8cm,width=9cm]{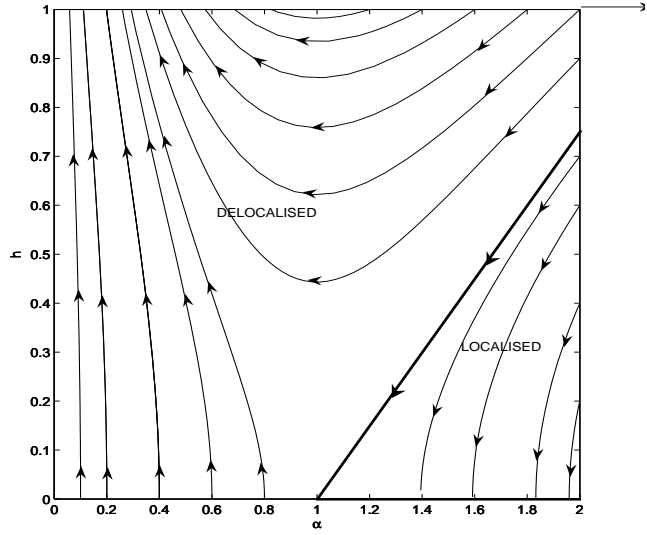}}\qquad
\caption{RG flow for the ohmic SBM ($s=1$). Here, the flow is constructed 
numerically by giving various initial (bare) values of the coupling constants $h$ 
and $\alpha$. See discussion in Sec. \ref{subsec:ohmic} for the interpretation.}
%\label{ja:fig:1}
\label{se:fig:ohmic}
\end{figure} 

 The situation of the ohmic bath spectrum is probably one of the most well
understood case in the study of SBM. This is due to the fact that a linear
dispersion of the bath DoS lends itself to an exact mapping to the anisotropic
Kondo Model (AKM) \cite{theo,legget}. Due to this mapping, it is known that 
the critical dissipation occurs at $\alpha_c = 1$ for small non-zero $\Delta$,
and that the phase transition is of the Kosterlitz-Thouless type (infinite
order).
The flow equations that are found through the mapping to the AKM match the 
$\beta$ functions that we obtained by using the Majorana representation (see 
Eq.~\ref{eq:rg4} and Eq.~\ref{eq:rg5}). From the structure of these $\beta$
functions of the ohmic SBM, it is amply clear that the term $-\alpha h^2$ drives the 
dissipative coupling to zero whenever $\alpha<1$. However, in the regime $\alpha>1$,
it is now the transverse field term $h$ that is driven to zero, with the dissipative 
coupling $\alpha$ renormalizing to a finite value. 
Furthermore, in the limit $\alpha>1$ we see that the RG equations, Eq.~\ref{eq:rg4} 
and Eq.~\ref{eq:rg5}, have in fact a line of stable fixed points at zero field,
the typical signature of a phase transition of the Kosterlitz-Thouless type. 
This discussion is encapsulated by Fig.~\ref{se:fig:ohmic}, which represents 
the various RG trajectories that are obtained by numerically solving Eq.~\ref{eq:rg4} 
and Eq.~\ref{eq:rg5}, for various initial values of $\alpha$ and $h$. From this
flow diagram, it is clear that there exists a separatrix such that for any value 
of $\alpha$ and  $h$ that lies below the separatrix, the RG flow terminates at the 
line of fixed points, whereas if one were to start with initial value of $\alpha$ 
and $h$ lying above the separatrix, the flow maintains the system in the de-localized 
phase.

\subsection{The RG equations for the super-ohmic case ($s>1$)}
 \label{subsec:super}

 \begin{figure}
\centering
{\includegraphics[height=8cm,width=9cm]{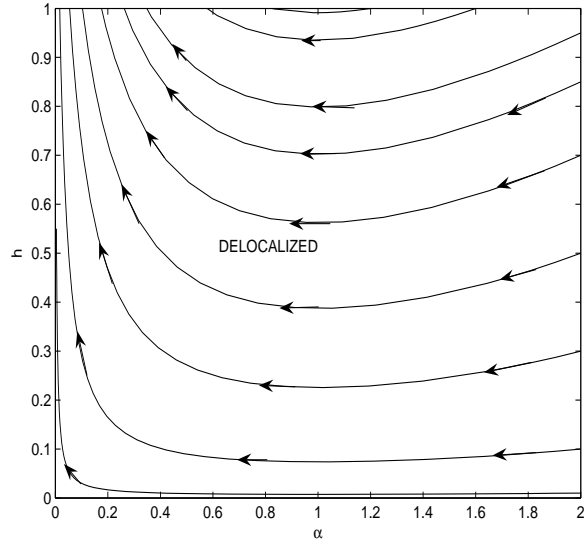}}\qquad
\caption{ Flow equations of the super-ohmic SBM model ($s>1$), starting with various 
initial values of $\alpha$ and $h$. For further details refer to Sec.~\ref{subsec:super} in the text.}
%\label{se:fig:sub}
\label{se:fig:super}
\end{figure} 

 In the situation where the bath spectrum is super-ohmic, i.e. $s>1$, it can
be readily argued that the system supports no critical fixed points. This fact
can be essentially gleaned from solving the set of equations, Eq.~\ref{eq:rg6}
and Eq.~\ref{eq:rg5} numerically for various initial configurations of $\alpha$
and $h$, giving the results depicted in Fig.~\ref{se:fig:super}. 
One sees that for any initial value of the dissipation and the transverse field, 
the couplings always flow towards the limit $h=\infty$ and $\alpha=0$. This implies that 
for the super-ohmic case one always ends up in the de-localized phase, and no
quantum phase trantion is allowed.

 \subsection{The RG equations for the sub-ohmic case ($0<s<1$)}
 \label{subsec:sub}

\begin{figure}
\centering
{\includegraphics[height=8cm,width=9cm]{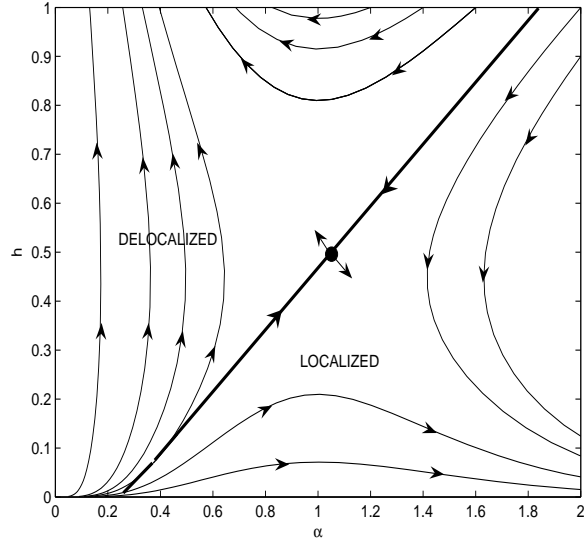}}\qquad
\caption{ Flow for the sub-ohmic SBM. Once again, the RG trajectories are plotted
for various initial values of the transverse field $h$ and the dissipative
coupling $\alpha$. For further details refer Sec.~\ref{subsec:sub} in the text.}
%\label{ja:fig:1}
\label{se:fig:sub}
\end{figure}

Now, we turn our attention to the most interesting case, namely the one where
the bath spectrum is endowed with a sub-ohmic dispersion, i.e. $0<s<1$. 
Since the mapping of the SBM to the AKM is invalidated in the
sub-ohmic regime, the situation of the SBM in the range $0<s<1$ was not fully
appreciated until recent study from Numerical Renormalization Group (NRG)
calculations \cite{bulla}, where a second order quantum phase transition was
explicitely demonstrated for all $0<s<1$. 
At this juncture, it should be noted that this localization/delocalization phase 
transition found in Ref.~\cite{bulla} is missed by the various other analytical
treatments of the sub-ohmic SBM, such as variational ansatz or diagonalizations
by unitary transformations, but is correctly predicted by the flow equations
derived above. The resulting flow is plotted in Fig.~\ref{se:fig:sub}, with a
fixed point occuring at $\alpha_c =1$ and $h=\sqrt{1-s}$, perturbatively controlled 
for values of $s$ close to 1.

 \section{Mapping to a long-ranged Ising Model}
 \label{sec:lri}
 In this section, we shall attempt to derive an effective model for the quantum
phase transition discussed previously, based purely in terms of the bosonic mode 
$\phi$. This can be done by representing the spin now in terms of 
Abrikosov fermions \cite{popov}, and then integrating the fermionic degrees of
freedom perturbatively in $\lambda$. By using this route we will see that the
SBM model can be mapped to a $\phi^4$ model with $O(1)$ symmetry and long-ranged
interactions in imaginary time. The mapping works as follows:
 \begin{equation}
 \vec{S} = \sum_{\s\s'} \fsd \frac{\vec{\sigma}_{\s\s'}}{2} \fsp 
 \label{eq:lri1}
 \end{equation}
Here, in Eq.~\ref{eq:lri1}, the $\fsd$ field are canonical fermions with an imaginary
chemical potential~\cite{popov}, that redefines the Matsubara frequencies
$\omega_n\to \omega_n+\pi T/4$,  and $\vec{\sigma}$ are the three Pauli matrices. 
By using Eq.~\ref{eq:lri1} to the defining Hamiltonian of the SBM, (Eq.~\ref{eq:1.1}) can 
be re-written in terms of the following action:
\begin{equation}
S = \int_{0}^{\beta} \! d\tau \, f^{\dagger} \left[ \partial_{\tau} - \frac{\Delta}{2}\sigma^x +
\frac{\lambda}{2}\sigma^z\sum_i(a_i+a_i^{\dagger})\right]f + 
\int_{0}^{\beta} \! d\tau \, \sum_i a_i^{\dagger}(\partial_{\tau} + \omega_i)a_i.
\label{eq:lri2}
\end{equation}
In the above equation $f$ is a two component vector whose Hermitian conjugate is
given by $f^{\dagger} \equiv (f^{\dagger}_{\uparrow}, f^{\dagger}_{\downarrow})$. 
Also in Eq.~\ref{eq:lri2}, the $\partial_{\tau}$ term is a consequence of time slicing 
when going into the path integral representation. The philosophy is now to formally 
integrate out the fermions to get an perturbative expansion in $\lambda$ of the
effective action. This methodology of integrating out the fermions is very similar 
in spirit to the treatment by Hertz of the itinerant ferromagnet \cite{hertz}, 
that we have already alluded to in the introductory section~\ref{sec:intro}. 
This technical step can be formally performed as the fermionic sector is purely
Gaussian, so that the effective theory reads:
\begin{equation}
S_{{\rm eff}} = \int_{0}^{\beta}\! d\tau\, \sum_i a_i^{\dagger}(\partial_{\tau} + \omega_i)a_i
- {\rm Tr} \ln \left[ \partial_{\tau} - \frac{\Delta}{2}\sigma^x +
\frac{\lambda}{2}\sigma^z\sum_i(a_i+a_i^{\dagger})\right].
\label{eq:lri3}
\end{equation}
Defining the ``local mode'' $\phi=\sum_i(a_i^\dagger+a_i)$, the bath can be
exactly encapsulated by the following Gaussian action, 
written with the Matsubara frequency $\nu_n=2n\pi T$:
\begin{equation}
S_{\rm eff}^{\rm Gauss} = -\frac{1}{\beta}\sum_{\nu_n}{\mathscr{G}}_{0}^{-1}(i\nu_n)
\phi(i\nu_n) \phi(-i\nu_n)
\label{eq:lri4}
\end{equation}
The quantity $\mathscr{G}_0$ in the above equation is given by
\begin{equation}
\mathscr{G}_0(i\nu_n) = \sum_i \left(\frac{1}{i\nu_n - \omega_i} +
\frac{1}{-i\nu_n-\omega_i}\right)
\label{eq:lri5}
\end{equation}
which can be re-expressed in terms of a spectral representation 
as:
\begin{equation}
\mathscr{G}_0(i\nu_n) = \int d\omega \frac{\rho(\w)}{i\nu_n - \w}.
\label{eq:lri6}
\end{equation}
Here, in Eq.~\ref{eq:lri6}, the bosonic density of states, $\rho(\w)$ is given by
\begin{equation}
\rho(\omega) = \frac{\w^s}{\w_c^{1+s}}\theta(\w_c+\w)\theta(\w_c-\w) {\rm{sgn}}(\omega).
\label{eq:lri7}
\end{equation}
The Green's function $\mathscr{G}_0$ can be calculated by substituting the
functional form of $\rho(\w)$ from Eq.~\ref{eq:lri7} into Eq.~\ref{eq:lri6}, and
then performing the integration over the frequency variable $\w$. Once we have
performed the integration, the resultant expression can be easily inverted to
obtain the functional form for ${\mathscr{G}}_{0}^{-1}$ which is given in the
low frequency limit by
\begin{equation}
\mathscr{{G}}_{0}^{-1}(i\nu_n) = -\frac{s \w_c}{2} - {\left(\frac{s \w_c}{2}\right)}^2
\frac{\pi\; |\nu_n|^s}{\sin\frac{\pi s}{2} \w_c^{s+1}}
\label{eq:lri8}
\end{equation}
Thus, substituting the above functional form of the ${\mathscr{G}}_{0}^{-1}$
into Eq.~\ref{eq:lri4}, we see that the Gaussian part of the action for the bath
of harmonic oscillators, Eq.~\ref{eq:lri3}, can be written as:
\begin{equation}
S_{\rm eff}^{\rm Gauss} = \frac{1}{\beta}\sum_{\nu_n}\left[\frac{s \w_c}{2} +
{\left(\frac{s \w_c}{2}\right)}^2
\frac{\pi\; |\nu_n|^s}{\sin\frac{\pi s}{2} \w_c^{s+1}}\right]|\phi(\nu_n)|^2
\label{eq:lri9}
\end{equation}

Now, that we have taken care of the Gaussian bath term in Eq.~\ref{eq:lri3}, we turn
our attention to the $Tr\ln$ term, which can be Taylor expanded to obtain:
\begin{equation}
\sum_{n=1}^{\infty}\frac{1}{n} Tr {\left(G_0 \sigma^z
\phi \frac{\lambda}{2}\right)}^{2n},
\label{eq:lri9b}
\end{equation}
wherein in the above equation, the quantity $G_0$ is endowed the functional
form, $G_0 = \frac{-i\w_n \mathbf{1}_2 +\frac{\Delta}{2}\sigma^x}{\w_n^2
+\frac{{\Delta}^2}{4}}$, where $\mathbf{1}_2$ is the usual $2\times 2$ identity
matrix. 
%Here, in the above equation
%and group the terms in such a fashion that we can 
%effectivel re-write Eq.~\ref{eq:lri3} as:
%\begin{equation}
%S_{\rm eff} = S_{\rm eff}^{\rm Gauss} + \sum_{n=2}^{\infty} a_n \sum_{\nu_1\ldots\nu_{n-1}}
%\chi(\nu_1\ldots\nu_{n-1})\phi(\nu_1)\ldots\phi(-\nu_1\ldots-\nu_{n-1})
%\label{eq:lri10}
%\end{equation}
Now, we proceed to calculate the traces implicit in Eq.~\ref{eq:lri9b}. The
resultant expression, up to order $\lambda^4$, is then combined with Eq.~\ref{eq:lri9}, 
to obtain:
\begin{equation}
S_{eff} = \int \frac{d\nu}{2\pi} \left(r+A|\nu|^s\right) |\phi(i\nu)|^2 +
\int d\tau \; u \;(\phi(\tau))^4.
\label{eq:Seff}
\end{equation}
Here $r = s\w_c/2 - \lambda^2/(4\Delta)$ is a mass term that controls the distance to 
criticality, $u=\lambda^4/(16\Delta^3)$ is the leading interaction term, and
$A\propto\w_c^{s-1}$. We note that this action is equivalent to an Ising model
in imaginary time, with interaction decaying as $1/(\tau-\tau')^{1+s}$, as
expected from the quantum/classical equivalence~\cite{emery,bulla}. Now, one
can use simple power-counting arguments to capture the critical behavior of the
long-ranged Ising model displayed in Eq.~\ref{eq:Seff}, thereby also
understanding the critical behavior of the underlying microscopic model,
Eq.~\ref{eq:1.1}. By doing a power counting analysis around the Gaussian fixed
point one finds that the scale dimension of the O$(\phi^4)$ term is $[u] =
2s-1$. This implies that the for all $s<1/2$ the scale dimension is negative
thus implying that the critical behavior is mean field like. However, for
$s>1/2$ one needs to account for higher loop effects to capture the true
critical behavior, leading to non trivial exponents with respect to the mean
field values.

\section{A special identity}
\label{sec:ward}
In this section, we will derive a special identity that helps us to calculate the
fully dressed bosonic propagator in terms of the spin-spin correlator
$\chi_z(\tau) = \langle\sigma^z(\tau)\sigma^z(0)\rangle$. We start
with an effective action which is a variant of the one that can be obtained from
Eq.~\ref{eq:1.1}. Thus, we have 
\begin{eqnarray}
S[\sigma, \phi, {\rm{J}}] = S_{\rm Berry} - \int d\T \frac{\Delta}{2} \sigma_z(\T) 
+ \frac{\lambda}{2}\int d\T \sigma_x \phi(\T) \nonumber\\
+\sum_\alpha \int d\T J^\alpha(\T)\sigma^\alpha(\T) + \int d\T d\T^{\prime}
\mathcal{G}_{0}^{-1}(\T-\T')\phi(\T)\phi(\T').
\label{eq:w1}
\end{eqnarray}
Here, in Eq.~\ref{eq:w1}, the term $S_{\rm Berry}$ is the so-called Berry action
that encodes the impurity spin commutation relations.
This term is not explicitly written down as its functional form relies on
spin-coherent states, the discussion of which is beyond the scope of this
manuscript. Also, in Eq.~\ref{eq:w1}, $J$ is a source term for the spin
dynamics, and $\mathcal{G}_0$ is again the bare bosonic propagator. The 
spin-spin correlator can be easily derived from Eq.~\ref{eq:w1}, by performing 
an appropriate functional differentiation of the partition function ${Z}$ 
with respect to the source field $J$. Thus, we have
\begin{equation}
\chi_z(\T) = \frac{1}{{Z}}\frac{\delta^2 Z}{\delta J(\T) \delta J(0)}\vert_{J =0}
\label{eq:w2}
\end{equation}
Now, in performing the technical calculations inherent in Eq.~\ref{eq:w2}, one can
re-express the bosonic field $\phi$ in terms of a new field $\tilde{\phi} =
\frac{J}{\lambda} + \phi$. In doing so we use the fact that the partition
function $Z$ remains invariant under such a redefinition of the $\phi$
field. Thus, re-expressing the partition function in terms of the $\tilde{\phi}$
fields and then performing the functional differentiation, we are led to the
following relation that connects $\chi_z(\tau)$ to the full bosonic Green's
function $\mathscr{G}_{\phi}(\T-\T')=\langle\phi(\T)\phi(\T')\rangle$ and the 
bare bosonic Green's function $\mathcal{G}_0$: 
 \begin{equation}
 \chi_z(\tau) = -\frac{4}{\lambda^2} \mathcal{G}_{0}^{-1}(\T) + \frac{4}{\lambda^2} \int d\T_1 d\T_2
\mathcal{G}_{0}^{-1}(\T_1-\T)\mathcal{G}_{0}^{-1}(\T_2)\langle \phi(\T_1)\phi(\T_2)\rangle
 \label{eq:w3} 
 \end{equation}
By going into the frequency domain representation we can compactly re-write Eq.~\ref{eq:w3} as
 \begin{equation}
 \chi_z(i\nu_n) = 
 -\frac{4}{\lambda^2} \frac{1}{\mathcal{G}_{0}(i\nu_n)}
+\frac{4}{\lambda^2} \frac{\mathscr{G}_{\phi}(i\nu_n)}{{\mathcal{G}_{0}(i\nu_n)}^2}
 \label{eq:w4}
 \end{equation}
This identity couples the single particle full bosonic propagator
$\mathscr{G}_{\phi}$ to the spin-spin susceptibility $\chi_z$, naively a four
operators correlation function (see e.g. the decomposition onto Abrikosov
fermions), and shows that both the bosonic field and the longitudinal spin density 
must become critical altogether at the quantum phase transition.
This formula becomes extremely useful in the context of diagrammatic
expansions that use the Majorana representation (introduced in
section~\ref{sec:maj}), because the spin susceptibility is simply related to
the single particle Majorana propagators, leading to a very powerful Ward
identity. These further theoretical developments go however much beyond
the scope of this review.
 
However, in the next section, Sec.~\ref{sec:shiba}, we shall show the usefulness of
the identity derived in this section to recover a well known result in
the context of SBM models, the so-called Shiba's relation.

\section{Shiba's Relation for the sub-ohmic spin boson model}
\label{sec:shiba} 
In this section we establish the Shiba's relation, usually discussed for
the ohmic SBM, in the case of the sub-ohmic model $s<1$.
This relation essentially connects the spin correlations at equilibrium to 
the zero-frequency spin susceptibility (via the free bath spectrum).
To derive this we use the fact that the exact bosonic Green's function,
on the real frequency axis, has the following low-frequency form (as can be checked 
by simple perturbative calculations from the effective action Eq. \ref{eq:Seff}) 
\begin{equation}
\mathcal{G}_{\phi}(\nu) = {(m+a_s^0{|\nu|}^s+i b_s^0{|\nu|}^s {\rm{sgn}}(\nu))}^{-1}
\label{s1}
\end{equation}
where $m$ is the renormalized mass driving the transition, and $a_s, b_s$ are
non critical numerical coefficients.
In the limit of small frequencies, the above reduces to:
\begin{equation}
\mathcal{G}_{\phi} (\nu) = \frac{1}{m} - i\frac{b_s^0}{m^2}{|\nu|}^s {\rm{sgn}}(\nu)
\label{s2}
\end{equation}
Now, the identity Eq.~\ref{eq:w4} gives us a relation that connects this
full bosonic Green's function $\mathcal{G}_{\phi}$ to the spin susceptibility.
At low frequency, and introducing the bare mass $m_0=1/\mathcal{G}_{0} (0)$,
one obviously gets:
\begin{equation}
\frac{1}{m}= \frac{1}{m_0} -\frac{{\lambda}^2}{4 m_0^2}{\chi_z^\prime}(0) 
\label{s3}
\end{equation}
and 
\begin{equation}
\chi_z^{\prime \prime} (\nu) = -b_s^0\frac{{\lambda}^2}{4 m_0^2}{|\nu|}^s
{\rm{sgn}}(\nu) {[\chi_{z}^\prime(0)]}^2
\label{s4}
\end{equation}
In Eq.~\ref{s3} and Eq.~\ref{s4}, the quantities $\chi_z^{\prime}$ and 
$\chi_z^{\prime \prime}$ are the real and imaginary part of the longitudinal
spin susceptibility $\chi_z$.
The imaginary part of the bare bosonic Green's function reads:
$\mathcal{G}_{0}^{\prime\prime}(\nu) = - \frac{J(|\nu|)}{{\lambda}^2} {\rm{sgn}}(\nu)
= -\frac{b_s^0}{{m_0}^2}{|\nu|}^s {\rm{sgn}}(\nu)$. 
Thus, substituting for $b_s^0$ in Eq.~\ref{s4}, we get:
\begin{equation}
\chi_z^{\prime \prime} (\nu) = \frac{1}{4}J(|\nu|) {\rm{sgn}}(\nu) {[\chi_z^\prime(0)]}^2
\label{s5}
\end{equation}
Finally, from the definition that at $T=0$ the imaginary part of the spin
susceptibility is related to the spin correlation function $C(\nu)$ via the simple
relation $\chi_z^{\prime \prime} = {\rm{sgn}}(\nu) C(\nu)$, we obtain
\begin{equation}
C(\nu) = \frac{1}{4}J(|\nu|) {[\chi_z^\prime(0)]}^2
\label{s6}
\end{equation}
which is the generalized Shiba relation for the sub-ohmic spin boson model, and
is valid in the low frequency limit for the whole delocalized phase.

\section{On the Quantum to Classical mapping}
\label{sec:qucl}

From the general arguments given in the introduction, and the detailed
derivation of the classical effective theory Eq.~\ref{eq:Seff} for
the specific case of the spin boson model, the results of 
Ref.~\cite{vojtaPRL} came as a surprise, since critical exponents associated
to the spin magnetization $\left<S^z\right>$ were numerically found by these authors to 
deviate from the expected mean field result for $0<s<1/2$. 
At the time of writing, this issue is still debated~\cite{winter}, see
Ref.~\cite{us_sbm} for a more recent update on the question.

\section*{Acknowledgements}
RN thanks Priyanka Mohan for generating some of the figures in this manuscript.
He also thanks her for discussions and valuable comments. 
%%%%%%%%%%%%%%%%%%%%%%%% referenc.tex %%%%%%%%%%%%%%%%%%%%%%%%%%%%%%
% sample references
% %
% Use this file as a template for your own input.
%
%%%%%%%%%%%%%%%%%%%%%%%% Springer-Verlag %%%%%%%%%%%%%%%%%%%%%%%%%%
%
% BibTeX users please use
% \bibliographystyle{}
% \bibliography{}

\begin{thebibliography}{99.}%
% Use the following syntax and markup for your references if 
% the subject of your book is from the field 
% "Mathematics, Physics, Statistics, Computer Science"
%
% Contribution 
\bibitem{belitz}D. Belitz, T.R. Kirkpatrick, T. Vojta, Rev. Mod. Phys. {\bf 77}, 579 (2005) 
\bibitem{bulla} R. Bulla, N-H Tong and M. Vojta, Phys. Rev. Lett. {\bf 91},
170601 (2003).
\bibitem{shraiman_giam} D. G. Clarke, T. Giamarchi and B. I. Shraiman,
Phys. Rev. B {\bf 48}, 7070 (1993). 
\bibitem{theo} T. A. Costi, and G. Zarand, Phys. Rev. B, {\bf 59}, 12398, (1999).
\bibitem{emery} V. J. Emery and A. Luther, Phys. Rev. B {\bf 9}, 215 (1974).
\bibitem{us_sbm} S. Florens, A. Freyn, D. Venturelli, and R. Narayanan, unpublished.
\bibitem{hertz} J. A. Hertz, Phys. Rev. B {\bf 14}, 1165 (1976).
\bibitem{larkin} A. I. Larkin and V. I. Mel'nikov, Sov. Phys. JETP {\bf 34},
656 (1972).
\bibitem{legget} A. J. Leggett, S. Chakravarty, A. T. Dorsey, M. P. A. Fisher,
A. Garg and W. Zwerger, Rev. Mod. Phys. {\bf 59}, 1 (1987).
\bibitem{loh} Y. L. Loh, V. Tripathi and M. Turlakov, Phys. Rev. B {\bf 71},
024429 (2005).
\bibitem{rmpvojta} H. v. L\"ohneysen, A. Rosch, M. Vojta and P. W\"olfle, Rev. Mod. Phys. {\bf 79}, 1015 (2007) 
\bibitem{mahan} G. D. Mahan, Many-Particle Physics, Second Edition, Plenum Press, New York, (1990).
\bibitem{novais} E. Novais, A. H. Castro Neto, L. Borda, I. Affleck and G.
Zarand, Phys. Rev. B {\bf 72}, 014417 (2005).
\bibitem{popov} V. N. Popov and S. A. Fedotov, Sov. Phys. JETP {\bf 67}, 535 (1988).
\bibitem{sengupta} A. M. Sengupta, Phys. Rev. B {\bf 61}, 4041 (2000).
\bibitem{georges} A. M. Sengupta and A. Georges, Phys. Rev. B {\bf 52}, 10295
(1995).
\bibitem{si} Q. Si and J. Lleweilun Smith, Phys. Rev. Lett. {\bf 77}, 3339
(1996).
\bibitem{vojta_bura} M. Vojta, C. Buragohain, and S. Sachdev,
Phys. Rev. B {\bf 61}, 15152 (2000).


% Online Document

\bibitem{davide} D. Venturelli, Masters thesis, un-published.
\bibitem{millis} A. J. Millis, Phys. Rev. B {\bf 48}, 7183 (1993).
\bibitem{mao} W. Mao, P. Coleman, C. Hooley and D. Langreth, Phys. Rev. Lett. 
{\bf 91}, 207203 (2003).
\bibitem{shnir} A. Shnirman and Y. Makhlin, Phys. Rev. Lett. {\bf 91}, 207204 (2003).
\bibitem{sachdev} S. Sachdev, Phys. Rev. B {\bf 55}, 142 (1997)
\bibitem{vojtaPRL} M. Vojta, N. Tong, and R. Bulla, Phys. Rev. Lett. {\bf 94},
070604 (2005).
\bibitem{winter} A. Winter {\it et al.}, Phys. Rev. Lett. {\bf 102}, 030601 (2009).
\end{thebibliography}
%

\end{document}